# EEG-based performance-driven adaptive automated hazard alerting system in security surveillance support

Xiaoshan Zhou, Pin-Chao Liao*

Department of Construction Management, Tsinghua University, No. 30, Shuangqing Rd., HaiDian District, Beijing 100084, PR China.

Xiaoshan Zhou: zhouxs20@mails.tsinghua.edu.cn (email address), (+86)18801136521 (Tel number)

Pin-Chao Liao (*Corresponding author): pinchao@tsinghua.edu.cn (email address), (+86)18618268224 (Tel number)

# Abstract

Computer-vision technologies have emerged to assist security surveillance. However, automation alert/alarm systems often apply a low-beta threshold to avoid misses and generates excessive false alarms. This study proposed an adaptive hazard diagnosis and alarm system with adjustable alert threshold levels based on environmental scenarios and operators' hazard recognition performance. We recorded electroencephalogram (EEG) data during hazard recognition tasks. The linear ballistic accumulator model was used to decompose the response time into several psychological subcomponents, which were further estimated by a Markov chain Monte Carlo algorithm and compared among different types of hazardous scenarios. Participants were most cautious about falling hazards, followed by electricity hazards, and had the least conservative attitude toward structural hazards. Participants were classified into three performance-level subgroups using a latent profile analysis based on task accuracy. We applied the transfer learning paradigm to classify subgroups based on their time-frequency representations of EEG data. Additionally, two continual learning strategies were investigated to ensure a robust adaptation of the model to predict participants' performance levels in different hazardous scenarios. These findings can be leveraged in real-world brain-computer interface applications, which will provide human trust in automation and promote the successful implementation of alarm technologies.

# 1 Introduction

Safety surveillance is critical for proactive management in construction to mitigate possible hazards ahead of time (Spencer, Hoskere, & Narazaki, 2019). Several security surveillance systems relying on closed-circuit television (Hodgetts, Vachon, Chamberland, & Tremblay, 2017), stereo vision cameras (Son & Kim, 2010), unmanned aerial vehicles (D. Kim, Liu, Lee, & Kamat, 2019; Roberts, Bretl, & Golparvar-Fard, 2017), and LiDAR (Gargoum & Karsten, 2021) have been developed, providing remote solutions for project monitoring. Operators who monitor these captured image data for hazard inspection are however faced with several inherent cognitive challenges, such as bias (Hinze, Godfrey, & Rinker, 2003), fatigue (Marcora, Staiano, & Manning, 2009), complacency (Haslam et al., 2005), stress (Leung, Chan, & Olomolaiye, 2008), and distractions (Chen, Song, & Lin, 2016). As a result, up to 50% of hazards are reported to be unrecognized during manual observation (Hardison & Gray, 2021; Sun & Liao, 2019), raising concerns for both practitioners and researchers in the construction industry.

To address the limitations of current manual efforts, various state-of-the-art computer vision (CV) technologies that create situation assessments enabling diagnosis, reasoning, and decision support have emerged (W. Liu, Meng, Li, & Hu, 2021; Mostafa & Hegazy, 2021; Paneru & Jeelani, 2021). Although advanced automation is now being developed and continues to become increasingly autonomous, the value of automated systems resides not in their total replacement of human operators but rather in their ability to augment the operators' capacities (Schaefer, Chen, Szalma, & Hancock, 2016). Hybrid human-machine collaboration (HMC) systems in hazard recognition tasks have thus been developed, bringing new concerns related to human trust in automation (Y. Liu & Jebelli, 2022b; You, Kim, Lee, Kamat, & Robert, 2018).

For the design of automated diagnosis and alarm systems, a critical challenge is determining the alert threshold. Most alarm systems have a low beta threshold because the costs of misses (e.g., workers' injuries and fatalities, structure collapse, and electric leakage) are typically much greater than the costs of false alarms. However, a predictable issue is that this causes most alerts to be "false alarms," which has two negative consequences. First, because it is the human who makes the final decision, frequent false alarms will make the human cross-check the raw data to ensure that the alert is indeed false, which may create distraction and lead to the expenditure of unneeded efforts (Okpala, Parajuli, Nnaji, & Awolusi, 2020). Second and more seriously, after excessive false alarms, people may develop a "cry wolf" syndrome that might result in them responding to alerts (including those that may be true) late or ignoring them altogether (Wickens et al., 2009), which may further cause catastrophic system failure and fatal accidents (Merritt, Heimbaugh, LaChapell, & Lee, 2013; N. Stanton & Walker, 2011; Woods, 2019). With the development of multiple wearable sensing devices (WSDs) as ergonomic tools for physiological monitoring to provide early warnings (Awolusi, Marks, & Hallowell, 2018), the problem is amplified when an operator receives alerts from several different independent systems. Therefore, automation must not only go beyond acting as a human backup but also support the process of human trust (Chiou & Lee, 2021; Shayan Shayesteh & Jebelli, 2022).

Previous research has shown that multiple alert threshold settings in automated hazard diagnosis systems can improve the HMC experience (Zhong, 2021). Thus, to address the problem of excessive false alarms and improve human trust in automation, we proposed an adaptive automated diagnostic tool with alert thresholds that may change according to environmental scenarios and human operators' performance during system operations to improve on-site security surveillance. To enable direct communication between humans and automated systems, brain-computer interface (BCI) technology may be used (Y. Liu, Habibnezhad, & Jebelli, 2021a). A BCI enables nonverbal communication between humans and automated systems by directly decoding human intentions based on brain activity patterns (Y. Liu, Habibnezhad, & Jebelli, 2021a; Zhou, Hu, Liao, & Zhang, 2021), which have been transformed into commands to control a robot (Y. Liu & Jebelli, 2022a).

Inspired by the intuitive communication enabled by BCI in HMC, this study sought to explore the potential of constructing a BCI-based adaptive hazard diagnosis and alarm system with alerting thresholds that are adaptive to the hazard recognition performance of human operators inferred from the operators' brain activities captured by electroencephalogram (EEG) signals. Furthermore, to ensure that the joint system can work robustly in the real world, the BCI must exhibit learning capabilities that enable it to respond to changing hazardous scenarios; in other words, the BCI needs to adapt the accumulated skills acquired from previous scenarios to new scenarios. Thus, this study also investigated the effectiveness of continual learning techniques in improving the plasticity of the prediction model among the scenarios. The main contributions of this study are as follows:

- A detailed comparative study was carried out for EEG-enabled adaptive systems in the construction field. The results revealed that a performance-driven adaptive system can work more robustly than a commonly proposed workload-driven system for HMC hazard inspection.

- To examine the feasibility of the proposed EEG-based performance-driven adaptive aiding system, a case study was conducted to verify the hypothesis that an EEG-based BCI can reliably distinguish the brain activation patterns elicited by participants with high, medium, and low hazard recognition performance levels.

- To the best of our knowledge, this study is the first to investigate how continual learning can improve skill transfer between scenarios in EEG-based BCI systems.

The remainder of this paper is organized as follows. Section 2 describes previous research on the concept and implementation of adaptive automation. Section 3 introduces the experimental setup and methodology. Section 4 presents and discusses the results of the data analysis. Finally, Section 5 outlines the conclusions and identifies directions for future research.

## 2 Related Work

**2.1 Conceptualizing adaptive automation**

To promote human trust and calibrate appropriate reliance on automation, two conceptually attractive strategies—adaptable/adaptive automation—have been proposed in the field of HMC (de Winter & Dodou, 2014). The first emphasizes on giving human operators the choice to invoke or remove higher levels of automation by self-monitoring their capacity of performing tasks, which helps operators better adapt to the functional characteristics of the working system (Ferris, Sarter, & Wickens, 2010). Following this idea, appropriate operator selection and training are needed to improve human understanding of the automation working principle (Chavaillaz & Sauer, 2016). However, a prominent relevant concern is the tendency for humans to be overconfident and inaccurate in subjective estimates of their performance and workload (Horrey, Lesch, & Garabet, 2009). It was also reported that this adaptable approach could only improve the overall system performance to a limited extent (de Winter & Hancock, 2021). Therefore, it is suggested that "For a viable future, technology must adapt to the human, which underwrites the necessity of human factors science" (de Winter & Hancock, 2021). Adaptive automation, proposed in the 1990s, provides a new form of HMC by allowing automated systems to automatically adapt to humans as a function of the environmental state, human state, or task performance (Byrne & Parasuraman, 1996; Scerbo, 1996).

**2.2 Workload-driven vs. performance-driven adaptive automation**

In recent years, various EEG-based WSDs have been developed in the construction field to track workers' health-related physiological data indicating fatigue (Aryal, Ghahramani, & Becerik-Gerber, 2017), vigilance and attention (Wang et al., 2017), mental workload (Chen, Song, & Lin, 2016), stress (H. Jebelli, Hwang, & Lee, 2018b), and emotions (Hwang, Jebelli, Choi, Choi, & Lee, 2018), and provide early warning signs of safety issues to construction workers to mitigate health risks and safety hazards on construction sites (Awolusi, Nnaji, Marks, & Hallowell, 2019; Tsao, Li, & Ma, 2019). Further, mental state monitoring of workers has been used to construct adaptive joint HMC systems that include a wearable biosensor for assessing workers' psychological conditions such that the automated system can adjust its working style accordingly to facilitate human trust in automation (Y. Liu, Habibnezhad, Jebelli, & Monga, 2022; Shayan Shayesteh & Jebelli, 2022). In an adaptive alerting system, the alert threshold levels should be automatically adjusted according to the status of the operators. More specifically, when the operator is detected as working in a suboptimal state (e.g., high workload), adaptive aiding is triggered, which is also known as a "human-in-the-loop" approach in previous literature (Eskandar, Wang, & Razavi, 2020). Presently, a series of auxiliary systems has been developed based on real-time EEG-based workload measurements (Y. Liu & Jebelli, 2022b). Liu et.al have developed a brainwave-driven HMC paradigm, in which robots continuously monitor workers' cognitive load and react accordingly (Y. Liu, Habibnezhad, & Jebelli, 2021b). Compared to a manual condition or one where adaptive aiding is provided randomly, psychophysiological adaptive automation of tracking workload has proven to lead to a significant improvement in targeting performance (Shayan Shayesteh & Jebelli, 2022).

Although previous studies have appropriately characterized EEG features for workload driven HMC systems, some studies have also suggested that automated systems may not be embraced by several construction workers. For example, Shayesteh and Jebelli found that, instead of reducing, participants' cognitive load increases when collaborating with an autonomous robot (S. Shayesteh & Jebelli, 2023). Previous studies also indicated a trade-off dilemma between cognitive workload and situational awareness; two safety-critical variables for human operators are significantly influenced when the operators use automated aiding systems (Rusnock & Geiger, 2016). In other words, a demand in the task workload also reduces situational awareness (Heikoop, de Winter, van Arem, & Stanton, 2018). For instance, drivers reported that driving with adaptive cruise control (ACC) was less effortful than driving manually (Hoedemaeker & Brookhuis, 1998) and as a consequence, driving with ACC may reduce driver vigilance and increase driver distraction (N. A. Stanton, Young, & McCaulder, 1997). Such a dilemma between cognitive workload saving and operators' engagement elicitation results in workload-driven adaptive alerts that are less reliable in security surveillance support.

As the relationship between cognitive state and hazard recognition performance is still vague (Hancock & Matthews, 2018) and the purpose of such a joint system is to improve the overall hazard recognition performance, this study proposed implementing a direct performance-driven adaptive automated alert system, wherein alert threshold levels automatically decrease on inferring that the human operator cannot effectively recognize the hazardous situation (Inagaki, 2008). Research has found that this performance-driven approach can achieve a good balance between cognitive workload and situational awareness. For example, Parasuraman et al. used the performance on a change detection task to drive adaptive aiding (automatic target recognition, ATR). Compared to performance without the ATR or to static automation where the ATR was continuously available, the adaptive automation condition was associated with both reduced workload and increased situational awareness (Parasuraman, Cosenzo, & de Visser, 2009).

Previously, without real-time physiological measures, this idea of performance-driven adaptive automation was impossible in the absence of any overt behavioral output and ground-truth during hazard inspection in practice. In recent years, several studies have explored the possibility of using EEG to predict future cognitive performance (Ayaz, Curtin, Mark, Kraft, & Ziegler, 2019; Stikic et al., 2011).

This leads to the hypothesis that an EEG-enabled BCI can be established to predict construction hazard recognition performance; however, there is no direct evidence of whether EEG signals from individuals with high, medium, and low hazard recognition performance levels can be differentiated with classification accuracy exceeding the change level. Therefore, further research is required to investigate the potential feasibility of applying an EEG-based BCI to furnish a performance-driven adaptive aiding system for hazard inspection.

**2.3 Environmentally determined and continual learning**

Suitable alert thresholds for automated diagnostic systems vary significantly depending on external hazardous conditions. Selecting an appropriate threshold involves making a trade-off between miss and false alarm rates (Molloy, Ford, & Mejias, 2017). Although misses and excessive false alarms both degrade trust and adversely affect performance (Yamada & Kuchar, 2006), further research suggested that the degree of difficulty of the task, rather than the type of error, appears to influence the trust level. Trust has been found to degrade, particularly when automation misses or provides a false alarm while detecting a target that the operator perceives to be easily identifiable. However, trust in automation increases when the target is perceived to be difficult to identify (Madhavan & Wiegmann, 2007). In addition, alarms should be prompted to induce timely and consistent hazard avoidance actions. Late alarm timing has the potential to impair human trust in automation because of a conflict between expectation and alarm performance (Abe & Richardson, 2005) and inappropriate activation timing of hazard recognition schemas (N. Stanton & Walker, 2011). Thus, to improve trust in automated diagnostic systems, it is highly desirable to investigate how human operators differ in terms of perceived difficulty and response time to different types of construction hazards.

Moreover, for a BCI-enabled joint system in hazard inspection, we must consider the fact that the human brain activates differently in response to different hazardous situations (Jeon & Cai, 2021; Zhou, Hu, Liao, & Zhang, 2021). Although several studies have been conducted on BCI-enabled HMC communication, these studies were designed to achieve the best performance in a pre-designed task setting. As a result, although state-of-the-art performance has been achieved in these BCI systems, prediction models, typically neural networks, are incapable of adapting to changing environments. Known as catastrophic forgetting, standard neural networks forget most of the knowledge learned from previous tasks after training on subsequent tasks (French, 2006; McCloskey & Cohen, 1989). For a real-world adaptive diagnosis aiding system, static models cannot perform robustly because changing scenarios are presented over time. Basically, the models need to maintain a balance between plasticity (the ability to adapt to new knowledge) and stability (the ability to retain prior knowledge) (Biesialska, Biesialska, & Costa-jussa, 2020; Parisi & Lomonaco, 2020). The stability-plasticity dilemma is a phenomenon where extreme stability makes it difficult to learn sequential tasks, whereas excessive plasticity can cause forgetting of the previously learned information (Grossberg, 1980, 1982). To address the problem of a distribution shift, a possible solution is to repeat the training process using an extended dataset that involves both previous and current data. Nevertheless, previous studies have shown that repeated training on a larger dataset is computationally intensive. Thus, this study proposed an exploration of resource-efficient continual learning techniques in the proposed real-world adaptive aiding system.

## 3 Methodology

### 3.1 Participants

Seventy-six construction workers with normal or corrected-to-normal vision participated in the experiment. One participant was excluded from the analysis because his EEG signals contained excessive artifacts. Five participants were considered unreliable in the validation test (see "Stimuli and Experimental Protocol" for details). The final sample consisted of 70 male participants (21–60 years old; mean age 42.2 years, all Chinese). All participants signed an informed consent form before participation and received 100 RMB as monetary compensation. This study was approved by the Department of Civil Engineering of Tsinghua University.

**3.2 Stimuli and Experimental Protocol**

All stimuli (images) in this study were retrieved from an in-use construction safety management platform, which is a repository for safety reports from numerous projects (Xu, Chong, & Liao, 2019). During hazard inspection, the safety inspectors upload pictorial recordings and text descriptions of hazardous construction scenes to the platform; subsequently, the general contractors are notified to rectify the scene to ensure that it is safe. After rectification, pictorial recordings of the corresponding corrected scenes are uploaded. The experiment comprised 60 pairs of construction scenes, with each pair having two opposite conditions (hazardous or safe). Thus, 120 trials were conducted.

The participants were required to complete a hazard recognition task in this experiment (Zhou, Liao, & Xu, 2022). They viewed images of real-world construction scenes displayed on a computer screen and judged whether they were hazardous or safe ( Figure 1a). The experiment consisted of three sections: practice, formal, and validation sections. Prior to the official experiment, the participants were instructed on the experimental procedure and completed ten trials to familiarize themselves with the task. The stimuli used in the practice session were different from those used in the official session. In the official experiment, 120 images were presented in randomized order. To alleviate fatigue, a one-minute break was imposed every 30 trials, during which participants were instructed to sit back and relax with their eyes closed. Finally, a validation session was conducted in which participants responded to 30 trials randomly selected from the previous 120 trials. The consistency of the responses to the same stimulus was checked, and participants with an inconsistency rate above 50% were excluded from the data analysis. As a result, five participants were excluded.

The procedural details of each trial are shown in Figure 1b. Each trial began with a fixation cross that appeared for 500 ms. Thereafter, an image depicting a construction scene (e.g., a lift with an open door) was presented for a maximum of 3000 ms, followed by a blank screen for 500 ms. Subsequently, a response screen was shown during which the participant was required to report his judgement of the construction scene seen before by pressing the corresponding key on the keyboard ("0" for safe and "1" for hazardous). No time limit was set for response screening. On average, the official experimental session lasted approximately 14 min.

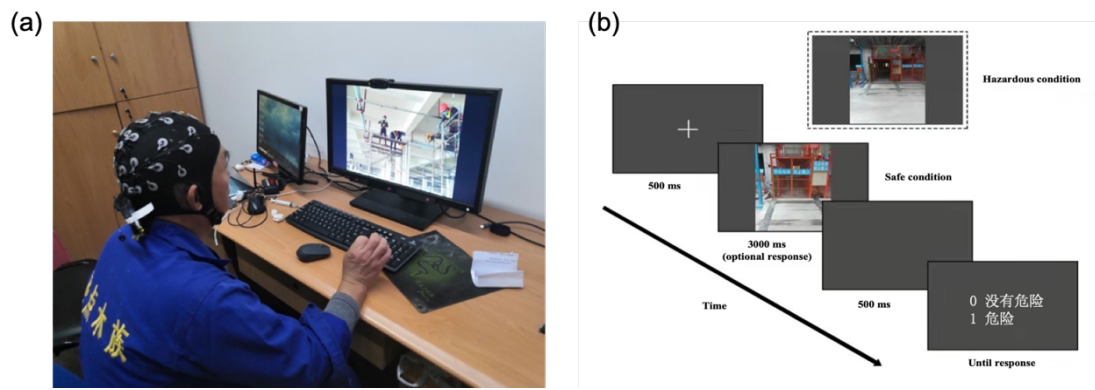

**Figure 2** (a) Experiment in which a participant is viewing the construction image with simultaneous recording of EEG signals. The details of the experimental paradigm are shown in (b).

**3.3 EEG Data Recording and Preprocessing**

A 32-channel electrode cap was used to record the EEG signals at a sampling rate of 250 Hz. The electrodes were placed in a 10–20 system. The impedance of each electrode was maintained at less than 20 kΩ during the experiment. Offline analysis of the EEG data was conducted using the FieldTrip toolbox (Oostenveld, Fries, Maris, & Schoffelen, 2011) in MATLAB (version R2019a, MathWorks, Inc., Natlick, MA, USA). The EEG signals were initially treated with band-pass filtering (0.1–40 Hz). Following previous EEG-based real-time system design in the construction field (Houtan Jebelli, Hwang, & Lee, 2018a), independent component analysis was conducted to remove the possible artifacts (heart rate, respiration responses, eye movements, etc.) from the EEG data. The corrected data of each trial were segmented into a [–200, 1000 ms] epoch (0 ms denotes the stimulus onset) with a 200-ms pre-stimulus baseline correction. Subsequently, epochs with values exceeding ±100 μv in any electrode were rejected to avoid possible artifact contamination.

### 3.4 Data Analysis

Figure 2 shows the flowchart of the study methodology. Table 1 describes the three types of hazards investigated, which are the most common types on construction sites (Fang et al., 2020).

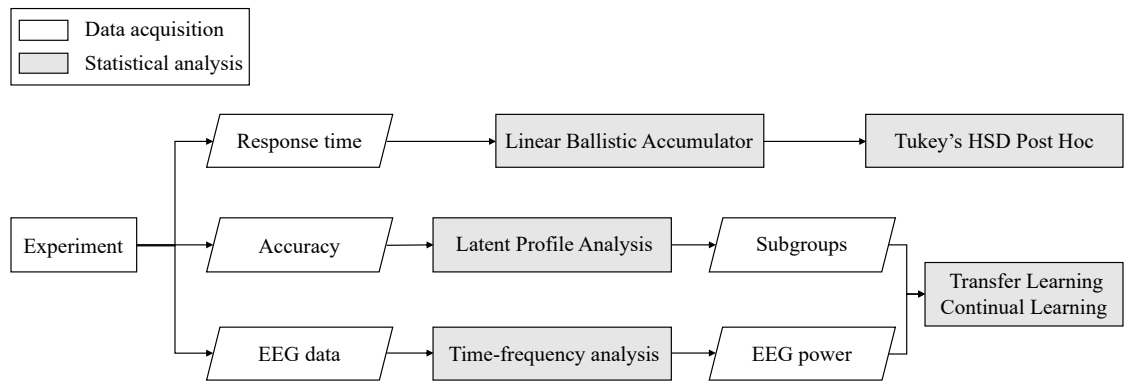

**Figure 2** Flowchart of the study methodology

**Table 1** Descriptions of the three hazard types

| Hazard Type | Descriptions of hazards | Example images of hazards | |
|---|---|---|---|
| | | Hazardous | Safe |

| Electric leakage | Overhead power lines; unprotected electrical panels; unclosed electrical compartment, etc. | 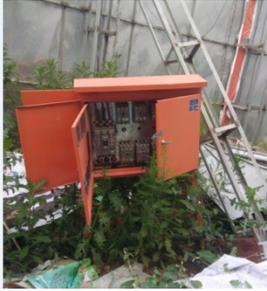 | 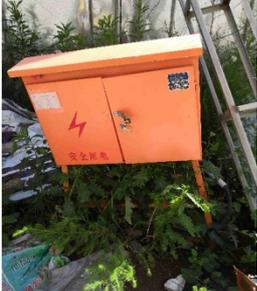 |
| --- | --- | --- | --- |
| Lack of edge protection | Slips, tris, and falls from a height (elevator hole, stair edge, suspended platform, etc.) without edge protection. | 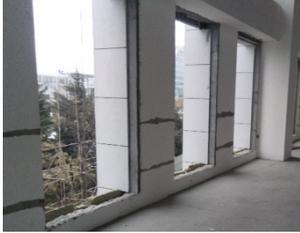 | 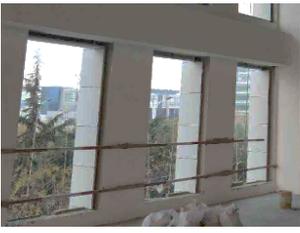 |
| Structural instability | Unstable temporal structures, such as scaffolding, formwork, and shoring. | 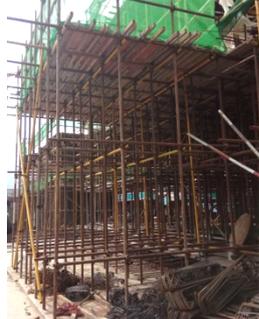 | 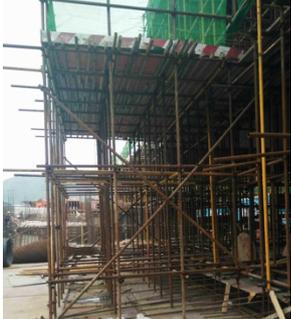 |

### 3.4.1 Linear Ballistic Accumulator Model

To gain deeper insight into the cognitive mechanisms of recognizing various hazards, the computational modeling of response time (RT), which allows the decomposition of RT into several different psychological functions involved in the hazard recognition process, could be conducted. Previous studies have applied the accumulated model to various cognitive processes (McIntosh & Sajda, 2020; Nishiguchi, Jiro, Kunisato, & Takano, 2019 & Takano, 2019). Although previous studies have typically indexed cognitive differences by mean differences in RT between various hazard types, RT contains a richer amount of information reflecting, for example, information processing efficiency and response conservativeness. To dissociate a single RT into different subprocesses involved in the hazard recognition response decision, we utilized a linear ballistic accumulator model (LBA) (S. Brown & Heathcote, 2005). In the LBA, response time is considered the time that participants take to accumulate evidence of hazardousness toward a decision threshold for characterizing the scene to be hazardous or safe. The hazard recognition process is described using four parameters (S. D. Brown & Heathcote, 2008): drift rate (*v*), representing the speed of evidence accumulation, i.e., efficiency in information processing; upper limit of the starting point distribution (*A*), representing the upper limit of the amount of existing evidence at the start of evidence accumulation (which varies across trials); threshold (*b*), representing the amount of evidence above which a "hazardous" decision is made; and non-decision time (*psi*), representing the time taken to encode stimuli and execute a response. With these parameters, the decision time is defined as the distance between the starting point and threshold divided by the drift rate (see Figure 3), whereas RT is composed of the decision and non-decision time (*psi*).

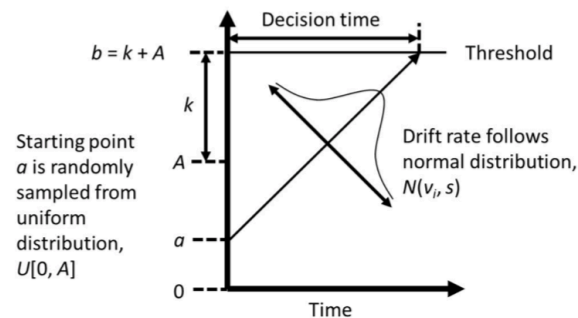

**Figure 3** Conceptual diagram of the linear ballistic accumulator model (Nishiguchi, Jiro, Kunisato, & Takano, 2019).

The LBA model parameters were estimated using the RSTAN package (Carpenter et al., 2017) written in R (version 4.0.2), which analyzes behavioral data using Bayesian inference. To avoid possible bias, only the trials with correct responses were included in the analysis. In the RSTAN package, the parameters were estimated using the Hamiltonian Monte Carlo algorithm (Neal, 2012), which uses the No-U-Turn sampler to sample posterior distributions with correlated parameters (Hoffman & Gelman, 2011). In this study, the Hamiltonian Monte Carlo algorithm (iteration = 4000, warmup = 2000, thinning = 1) was used to obtain the posterior distributions of each LBA parameter for each hazard type. Four Hamiltonian Monte Carlo chains were run to meet the Gelman–Rubin's criteria for convergence ((Gelman & Rubin, 1992), with R̂ close to 1).

**3.4.2 Latent Profile Analysis**

In this study, we used latent profile analysis (LPA) to segment participants into subgroups with three hazard recognition performance levels ("High", "Medium", and "Low"). LPA assumes that people can be categorized with varying degrees of probabilities into different configural profiles of personal attributes, which has received growing interest in occupational behavior research in recent years (Spurk, Hirschi, Wang, Valero, & Kauffeld, 2020). The *tidyLPA* R package (Rosenberg, Beymer, Anderson, & Schmidt, 2018) was used for the LPA analysis. The participants' recognition accuracies for various types of hazards were used as categorical latent variables. Our adopted model estimated the variances to be equal across profiles, while the covariances were constrained to zero. The LPA posterior probabilities were used to segment the participants into the corresponding subgroups. The subgroups were named based on the resultant probability of correct responses in all trials for each hazard type.

**3.4.3 Time-frequency Analysis of EEG signals**

The EEG signals were transformed into an image format after collection for the following classification analyses. Numerous studies have classified brain signals using traditional machine learning algorithms. Deep learning with convolutional neural networks (CNNs) has recently achieved outstanding performance in image classification. There is also increasing research interest in the use of CNNs for end-to-end EEG analysis (Schirrmeister et al., 2017); see (Craik, He, & Contreras-Vidal, 2019) for a comprehensive review. As deep learning has shown enormous potential in the computer vision field, advanced models, although complicated and inexplicable to some extent, provide state-of-the-art performance for image classification. In this study, we used the transfer learning approach to exploit some of the most prominent classification models in the literature. The rationale here was based on the improved EEG signal classification performances achieved by deep learning methods in previous studies (Li et al., 2017; Singh, Ahmed, Singh, Chanak, & Singh, 2020).

Following this idea, we converted EEG signals into time-frequency maps (Zhou, Liao, & Xu, 2022) using time-frequency analyses based on a wavelet transform approach (Tallon-Baudry, Bertrand, Delpuech, & Pernier, 1996). Analyses were performed using the FieldTrip toolbox (Oostenveld, Fries, Maris, & Schoffelen, 2011) in MATLAB. We used a Hanning taper with a 500 ms sliding window in 50 ms time steps to achieve a compromise between time and frequency resolutions. Each epoch was transformed in the frequency domain, and the time-frequency representation of the EEG signals was computed for each trial of each participant.

**3.4.4 Transfer learning for EEG signal classification**

As the EEG signals were transformed into images in the previous step, we applied the paradigm of transfer learning to conduct signal classification. The idea of transfer learning is as follows. For a classification problem, say $C_a$, we train the model on a set of data (say $D_s$). Now, for a different classification problem, say $C_b$, we do not have to train the model from scratch; rather, we can use the model trained on $D_s$ and apply the learned knowledge to the problem $C_b$. Thus, we take advantage of existing knowledge. In this study, we selected one of the most popular CNN models, ResNet18, for classification. The model, trained on approximately 1.2 million images from the ImageNet database (Deng et al., 2009), has shown impressive performance in several challenging circumstances. Following this, the ImageNet database became the source domain in which the model was trained, while the time-frequency maps of EEG signals were the target domain.

To train the ResNet18 model, we followed the customary approach described in the literature (Oquab, Bottou, Laptev, & Sivic, 2014). The top fully connected layer of ResNet18 was replaced with a new fully connected layer of 512 activation units, followed by a final layer with a LogSoftmax activation function to output the prediction. In the fine-tuning process, the early layers were kept fixed because they generate more generic features that can be used despite the data distribution; however, higher-level layers of the CNN may devote more representational power to features that are more specific to differentiating between categories. Therefore, we trained only the newly introduced layers using an Adam optimizer (Kingma & Ba, 2014) with a learning rate of 0.003. To assess the classification performance, we split the real data into train/validation subsets according to a split ratio of 80%:20%. To avoid overfitting, the training was stopped as soon as the loss on the validation set did not decrease in three epochs.

**3.4.5 Continual learning strategies**

To measure catastrophic forgetting, we first considered per-task baselines, i.e., the results of a model trained independently for each task. For continual learning, we considered the Naïve baseline, which was fine-tuned across tasks; the model was first trained on Task A and then on Task B, starting from the previously learned parameters. Fine-tuning assumes different tasks without much consideration of source performance, whereas continual learning does not consider forgetting the source domain and learning a target domain.

Continuous learning consists of two broad families of methods: rehearsal and regularization. The first assumes memory and access to explicit previous knowledge (instances), and the second only has access to compressed knowledge, such as previously learned parameters. In the rehearsal approach (Robins, 1995), the model is first trained on Task A; subsequently, the parameters are fine-tuned through batches taken from a dataset containing a small number of examples of Task A and the training set of Task B. Training examples for Task A were selected through uniform sampling. The main disadvantage of the rehearsal approach is that it requires a large storage capacity to preserve the raw samples or representations learned from the previous task. The regularization approach consolidates past knowledge using additional loss terms that slow down the learning of important weights used in the previously learned task. The most notable regularization-based approach is elastic weight consolidation (EWC) (Kirkpatrick et al., 2016).

In this study, the continual learning circumstance was domain-incremental learning, where the task structure remained consistent but the input distribution changed across sequential tasks. The task identity was unknown at the time of testing, and the model was only needed to solve the current task. This corresponds to a real-world adaptive aiding system that learns to operate in various scenarios without specifying that scenario.

## 4 Results and Discussion

**4.1 Behavioral Descriptive Statistics**

Figure 4 shows the mean accuracy and RTs for the two conditions of each hazard type. The differences across different types of hazards were checked using an analysis of variance (ANOVA) test. The results revealed that the accuracies of electricity and edge protection-related hazards were significantly higher than those of structural hazards (66.05% ± 9.87% for electricity, 64.36% ± 8.92% for edge protection, and 58.53% ± 12.18% for structure; $F = 10.04$, $p < 0.001$). There was no significant difference in RTs across the three hazard types ($F = 0.12$, $p = 0.88$).

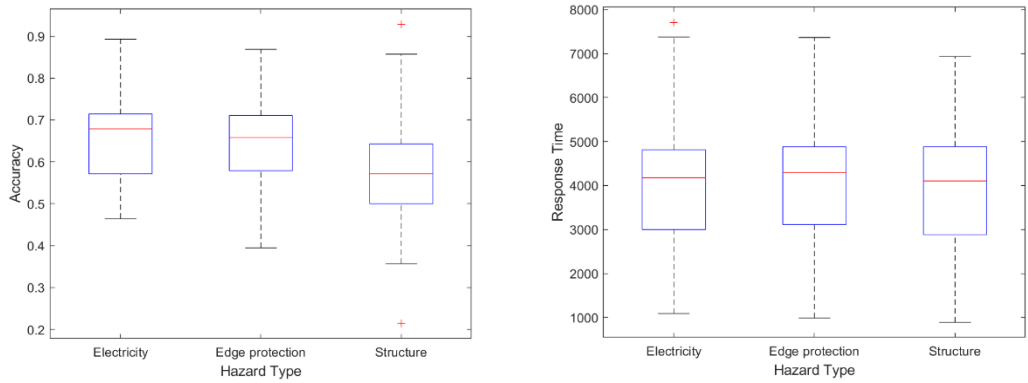

**Figure 4** Boxplots of accuracies and RTs (ms) by hazard type.

### 4.2 Analyses with LBA Modeling

To probe the cognitive differences in recognizing different types of hazards, 1714 correct trials for hazardous stimuli across all participants (hazard "Electric leakage" (EL): 640; hazard "Lack of Edge protection" (LEP): 781; hazard " Structural instability" (SI): 293) were submitted to LBA modeling. Satisfactory convergence was found for all estimated parameters according to Gelman–Rubin statistics: all R^ = 1.00. Figures 5 and 6 show the posterior distributions of these four parameters.

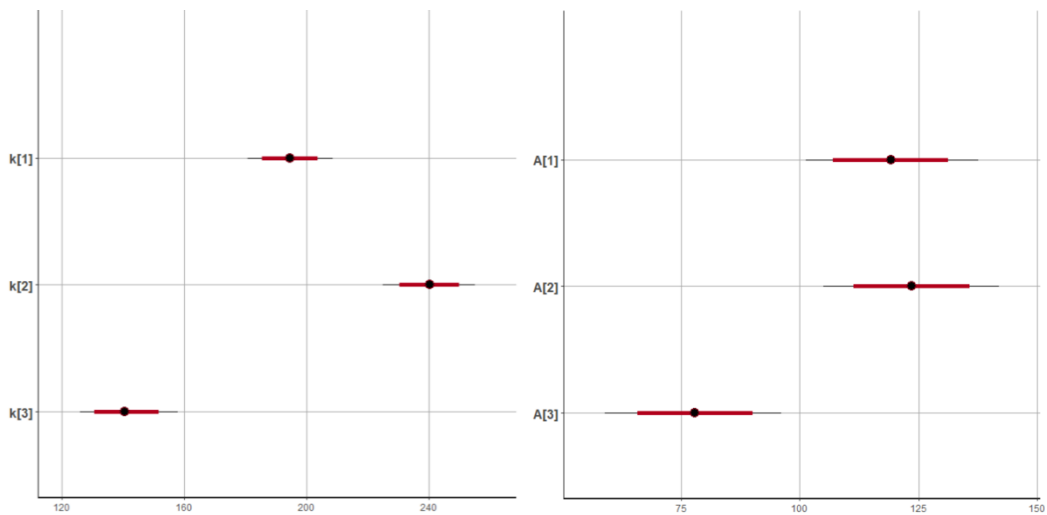

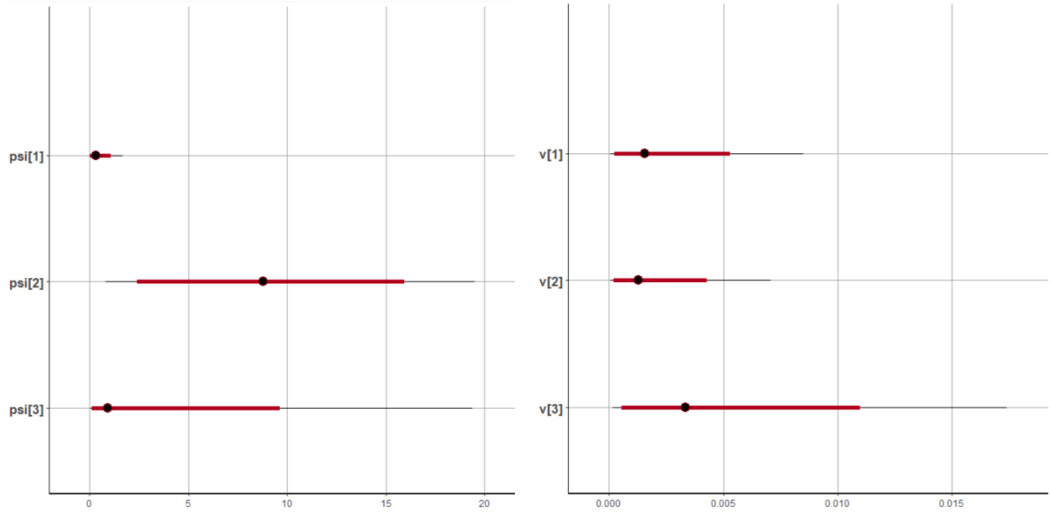

**Figure 5** Posterior uncertainty intervals ( 80% (inner, highlighted in red) and 95% (outer)) and the posterior median for all the parameters for correct hazardous trials by each hazard type.

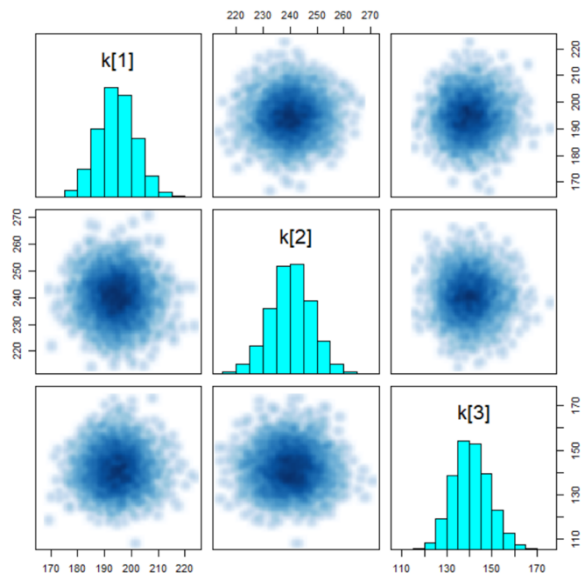

**Figure 6** Distributions of the relative threshold ($k$) for each of the three hazard types with histograms along the diagonal showing univariate marginal distributions, and scatterplots off the diagonal showing bivariate distributions.

The normality of the posterior distributions of the four parameters in each hazard type was tested using the Kolmogorov–Smirnov test (P > 0.05). Furthermore, the differences in parameters across the three types of hazards were checked using ANOVA, and significant differences were found in all four parameters (P < 0.001). Post hoc tests were conducted using Tukey's HSD, and significant differences were found in any pair of hazard types for each of the four parameters (p < 0.05).

Thus, although no significant differences were observed in RTs across different hazard types, further component decomposition of RT based on LBA revealed that participants adopted different sub-processes to correctly recognize various hazards. Recognizing LEP was accompanied by the highest decision threshold, which was reflected in the highest $k$ and $A$ parameters. Fall from height (FFH) accounts for most accidents and fatalities in construction (Nadhim, Hon, Xia, Stewart, & Fang, 2016) and remains a pervasive problem worldwide (Fang et al., 2019). The high tolerance for LEP demonstrated by the participants in this study provides deeper insight into the cause of FFH. The high decision threshold may also explain why previous studies reported the requirement of most cognitive resources to recognize falling hazards, as indicated by stronger brain activation (Liao, Sun, & Zhang, 2021); that is, participants are more prone to underestimate the risk of FFH than other types of hazards. To protect workers from FFH, installing barricades or edge protection is one of the most common practices. According to the Workplace Safety and Health Council, barricades are required for all building edges and edges of excavations, holes, floor openings, and roofs on construction sites (Workplace Safety and Health Council, 2013). Despite these policies, missing barricades remain a serious problem in construction. It has been found that lack of guardrails, handrails, barriers, and edge protection accounts for approximately one-third of fall-related accidents (Zlatar, Lago, Soares, Baptista, & Barkokébas Junior, 2019). Therefore, it is recommended that CV technologies are advanced particularly for the detection of missing barricades (Chian, Fang, Goh, & Tian, 2021; Kolar, Chen, & Luo, 2018) to mitigate falling risks. By contrast, participants were at least cautious in recognizing SI. Previous studies have shown that familiarity with tasks and cost of errors affect the decision threshold in recognition memory tasks (Andersen, Harrison, Lau, & Rutström, 2014; Juola, Caballero-Sanz, Muñoz-García, Botella, & Suero, 2019). Consequently, unfamiliarity (the evidence is not easy to obtain and is unclear) and serious consequences may lead to a low decision threshold to correctly recognize SI. Therefore, adaptive aiding for diagnosing structure-related hazards is emphasized to reduce manual efforts and potential human errors, such as advancements in Building Information Model-driven safety planning support for scaffolds (K. Kim & Cho, 2018) and temporary structures (K. Kim, Cho, & Kim, 2018).

**4.3 Analysis with LPA**

The LPA analysis separated participants into three performance-level subgroups. These were named "high", "medium", and "low," with the first subgroup having the highest probability of accurately recognizing all hazards and the last having the lowest probability of accurately performing the tasks (see Figure 7).

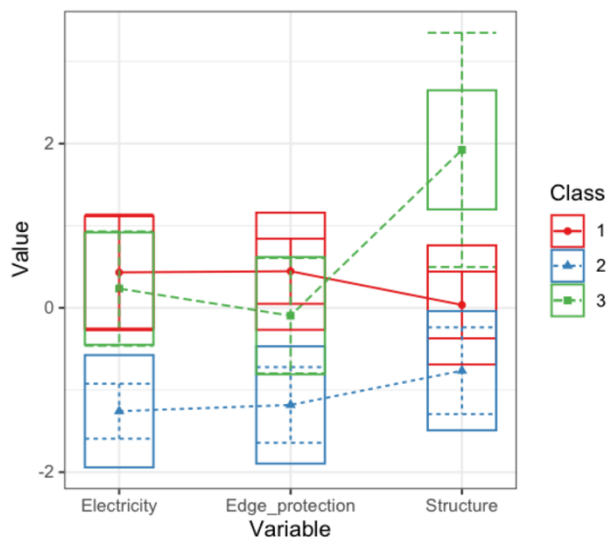

**Figure 7** Three profile classes of hazard recognition performance levels were identified amongst the participants: (1) high, (2) low, and (3) medium. Column centering and scaling were conducted to demonstrate the accuracy performance separation of each profile category. Centering involves subtracting column means from their corresponding columns, whereas scaling is conducted by dividing

columns by their standard deviations.

**4.4 EEG classification performance**

For practical applications, the BCI system should achieve a classification accuracy of at least chance level (=0.33). Table 2 summarizes the mean classification accuracies of all training strategies over the three runs. All task-specific models performed above the chance level (see the first row in Table 2). Although neural networks can provide good performance for individual tasks, learning multiple tasks sequentially remains a considerable challenge for deep learning. We observed a slight forgetting phenomenon between EL and LEP; the Naïve baseline, tested on LEP after fine-tuning on EL, achieved lower accuracy on the first task compared to being trained independently for both directions (0.687 vs. 0.699; 0.682 vs. 0.692). By contrast, SI introduced an asymmetric synergetic effect; EL or LEP exposure helped the model improve SI, achieving results that exceeded those obtained with the task-specific model (from 0.655 to 0.74 and 0.714, respectively); however, the effect was not symmetric as the accuracies on EL and LEP did not increase when SI was learned first.

Therefore, as the EEG-based BCI interacts with real-world environments that change the data distribution and catastrophic forgetting does exist as presented above, the BCI needs to be adaptive to the changing hazardous scenarios. Regarding whether continual learning helped to improve the model's plasticity, the results in Table 2 reveal that the order of tasks plays an important role. Although models that used continual learning strategies performed slightly better than the Naïve approach between EL and LEP in both directions, both EWC and rehearsal did not yield any improvement over the Naïve baseline when tested on SI after being fine-tuned on the other two hazards. However, after being exposed to SI, the model that used continual learning forgot less than the Naïve baseline in two setups: EWC and rehearsal performed better than the Naïve baseline on EL and LEP, respectively.

Table 2 Mean accuracy over three runs: trained on each task independently (first row: per-task) vs. sequential learning setups.

| ResNet18 | EL: 0.692 | | LEP: 0.699 | | SI: 0.655 | |
| --- | --- | --- | --- | --- | --- | --- |
| Sequential setups | EL→LEP | EL→SI | LEP→EL | LEP→SI | SI→EL | SI→LEP |
| Naïve | 0.687 | 0.74 | 0.682 | 0.714 | 0.677 | 0.684 |
| Rehearsal | 0.697 | 0.662 | 0.684 | 0.648 | 0.649 | 0.701 |
| EWC | 0.697 | 0.66 | 0.684 | 0.66 | 0.684 | 0.678 |

Note: EL denotes the hazard "Electric leakage", LEP denotes the hazard "Lack of edge protection", and SI denotes the hazard "Structural instability".

## 5 Conclusion and future work

As increasing numbers of CV- and WSD-based warning systems are developed, it is necessary to consider whether excessive warnings will adversely affect human hazard recognition performance owing to being a distraction and potentially cause distrust in automation. In this study, we proposed an automated diagnosis and alarm system with adjustable alert thresholds that is adaptive to hazard types and human operators' hazard recognition performance to promote the successful use of HMC for hazard inspection. A case study was designed and conducted to collect participants' accuracies, RT, and EEG signals while performing a construction hazard task. We first used LBA to decompose RT into several psychophysiological subcomponents, and the results revealed that participants demonstrated cognitive differences that were particularly reflected by the decision threshold while recognizing different hazard types. In addition, a preliminary analysis was conducted to investigate the feasibility of an EEG-based BCI system for acquiring real-time hazard recognition performance. We classified the EEG signals elicited by participants into high, medium, and low hazard recognition performance levels using the transfer learning paradigm. This analysis revealed that an EEG-based BCI is a highly promising solution for predicting task performance, with approximately 70% accuracy when trained independently on individual on-site scenarios. A distinctive contribution of this study is the proposal of a real-world BCI-based adaptive aiding system capable of continuously learning and adapting over time. To achieve this, we investigated how continual learning may help mitigate forgetting across scenarios. Two mainstream strategies for continual learning were explored, and the results revealed that the order in which models learn scenarios is important. For instance, SI→ EL facilitates the EWC strategy more than the other orders; however, there is no boost derived from exposure to EL first.

An important question for future research is why scenario order influences continual learning results. To answer this, we can explore how the human brain acts to prevent catastrophic forgetting and how CV algorithms can take inspiration from the human brain to implement continual learning among multiple tasks (van de Ven, Siegelmann, & Tolias, 2020). Moreover, algorithms should consider individual differences. In this study, the LBA model was fitted to trials across all participants to enable a large sample size; however, if a large number of observations were available, the parameters could be estimated for each individual participant. Furthermore, the algorithms can infer what has been acquired from the behavioral data to study the influence of scenario order on brain activation patterns to construct a more effective hierarchical BCI system for the BCI-enabled adaptive alerting aiding system.

## Funding

This work was supported by the National Natural Science Foundation of China [grant number 51878382].